\magnification 1200

%
%
\newdimen\FigSize       \FigSize=.9\hsize 
%
\newskip\abovefigskip   \newskip\belowfigskip
\gdef\epsfig#1;#2;{\par\vskip\abovefigskip\penalty -500
   {\everypar={}\epsfxsize=#1\nd
    \centerline{\epsfbox{#2}}}%
    \vskip\belowfigskip}%
%
\newskip\figtitleskip
\gdef\tepsfig#1;#2;#3{\par\vskip\abovefigskip\penalty -500
   {\everypar={}\epsfxsize=#1\nd
    \vbox
      {\centerline{\epsfbox{#2}}\vskip\figtitleskip
       \centerline{\figtitlefont#3}}}%
    \vskip\belowfigskip}%
%
\newcount\FigNr \global\FigNr=0
\gdef\nepsfig#1;#2;#3{\global\advance\FigNr by 1
   \tepsfig#1;#2;{Figure\space\the\FigNr.\space#3}}%
%
%
%
\gdef\ipsfig#1;#2;{
   \midinsert{\everypar={}\epsfxsize=#1\nd
              \centerline{\epsfbox{#2}}}%
   \endinsert}%
%
\gdef\tipsfig#1;#2;#3{\midinsert
   {\everypar={}\epsfxsize=#1\nd
    \vbox{\centerline{\epsfbox{#2}}%
          \vskip\figtitleskip
          \centerline{\figtitlefont#3}}}\endinsert}%
%
\gdef\nipsfig#1;#2;#3{\global\advance\FigNr by1%
  \tipsfig#1;#2;{Figure\space\the\FigNr.\space#3}}%
\newread\epsffilein    
\newif\ifepsffileok    
\newif\ifepsfbbfound   
\newif\ifepsfverbose   
\newdimen\epsfxsize    
\newdimen\epsfysize    
\newdimen\epsftsize    
\newdimen\epsfrsize    
\newdimen\epsftmp      
\newdimen\pspoints     
\pspoints=1bp          
\epsfxsize=0pt         
\epsfysize=0pt         
\def\epsfbox#1{\global\def\epsfllx{72}\global\def\epsflly{72}%
   \global\def\epsfurx{540}\global\def\epsfury{720}%
   \def\lbracket{[}\def\testit{#1}\ifx\testit\lbracket
   \let\next=\epsfgetlitbb\else\let\next=\epsfnormal\fi\next{#1}}%
\def\epsfgetlitbb#1#2 #3 #4 #5]#6{\epsfgrab #2 #3 #4 #5 .\\%
   \epsfsetgraph{#6}}%
\def\epsfnormal#1{\epsfgetbb{#1}\epsfsetgraph{#1}}%
\def\epsfgetbb#1{%
%
%
\openin\epsffilein=#1
\ifeof\epsffilein\errmessage{I couldn't open #1, will ignore it}\else
%
%
   {\epsffileoktrue \chardef\other=12
    \def\do##1{\catcode`##1=\other}\dospecials \catcode`\ =10
    \loop
       \read\epsffilein to \epsffileline
       \ifeof\epsffilein\epsffileokfalse\else
%
%
          \expandafter\epsfaux\epsffileline:. \\%
       \fi
   \ifepsffileok\repeat
   \ifepsfbbfound\else
    \ifepsfverbose\message{No bounding box comment in #1; using
defaults}\fi\fi
   }\closein\epsffilein\fi}%
%
%
\def\epsfsetgraph#1{%
   \epsfrsize=\epsfury\pspoints
   \advance\epsfrsize by-\epsflly\pspoints
   \epsftsize=\epsfurx\pspoints
   \advance\epsftsize by-\epsfllx\pspoints
%
%
   \epsfxsize\epsfsize\epsftsize\epsfrsize
   \ifnum\epsfxsize=0 \ifnum\epsfysize=0
      \epsfxsize=\epsftsize \epsfysize=\epsfrsize
%
%
     \else\epsftmp=\epsftsize \divide\epsftmp\epsfrsize
       \epsfxsize=\epsfysize \multiply\epsfxsize\epsftmp
       \multiply\epsftmp\epsfrsize \advance\epsftsize-\epsftmp
       \epsftmp=\epsfysize
       \loop \advance\epsftsize\epsftsize \divide\epsftmp 2
       \ifnum\epsftmp>0
          \ifnum\epsftsize<\epsfrsize\else
             \advance\epsftsize-\epsfrsize \advance\epsfxsize\epsftmp
\fi
       \repeat
     \fi
   \else\epsftmp=\epsfrsize \divide\epsftmp\epsftsize
     \epsfysize=\epsfxsize \multiply\epsfysize\epsftmp
     \multiply\epsftmp\epsftsize \advance\epsfrsize-\epsftmp
     \epsftmp=\epsfxsize
     \loop \advance\epsfrsize\epsfrsize \divide\epsftmp 2
     \ifnum\epsftmp>0
        \ifnum\epsfrsize<\epsftsize\else
           \advance\epsfrsize-\epsftsize \advance\epsfysize\epsftmp \fi
     \repeat
   \fi
%
%
   \ifepsfverbose\message{#1: width=\the\epsfxsize,
height=\the\epsfysize}\fi
   \epsftmp=10\epsfxsize \divide\epsftmp\pspoints
   \vbox to\epsfysize{\vfil\hbox to\epsfxsize{%
      \includegraphics{#1}%
      \hfil}}%
\epsfxsize=0pt\epsfysize=0pt}%
%
%
{\catcode`\%=12
\global\let\epsfpercent=
%
%
\long\def\epsfaux#1#2:#3\\{\ifx#1\epsfpercent
   \def\testit{#2}\ifx\testit\epsfbblit
      \epsfgrab #3 . . . \\%
      \epsffileokfalse
      \global\epsfbbfoundtrue
   \fi\else\ifx#1\par\else\epsffileokfalse\fi\fi}%
%
%
\def\epsfgrab #1 #2 #3 #4 #5\\{%
   \global\def\epsfllx{#1}\ifx\epsfllx\empty
      \epsfgrab #2 #3 #4 #5 .\\\else
   \global\def\epsflly{#2}%
   \global\def\epsfurx{#3}\global\def\epsfury{#4}\fi}%
%
%
\def\epsfsize#1#2{\epsfxsize}%
%
%

\epsfverbosetrue                        
\abovefigskip=\baselineskip             
\belowfigskip=\baselineskip             
\global\let\figtitlefont\bf             
\global\figtitleskip=.5\baselineskip    

\font\tenmsb=msbm10   
\font\sevenmsb=msbm7
\font\fivemsb=msbm5
\newfam\msbfam
\textfont\msbfam=\tenmsb
\scriptfont\msbfam=\sevenmsb
\scriptscriptfont\msbfam=\fivemsb
\def\Bbb#1{\fam\msbfam\relax#1}
\let\nd\noindent 

\def\natural{{\rm I\kern-.18em N}}
\newskip\ttglue


\def\eightpoint{\def\rm{\fam0\eightrm}  
  \textfont0=\eightrm \scriptfont0=\sixrm \scriptscriptfont0=\fiverm
  \textfont1=\eighti  \scriptfont1=\sixi  \scriptscriptfont1=\fivei
  \textfont2=\eightsy  \scriptfont2=\sixsy  \scriptscriptfont2=\fivesy
  \textfont3=\tenex  \scriptfont3=\tenex  \scriptscriptfont3=\tenex
  \textfont\itfam=\eightit  \def\it{\fam\itfam\eightit}
  \textfont\slfam=\eightsl  \def\sl{\fam\slfam\eightsl}
  \textfont\ttfam=\eighttt  \def\tt{\fam\ttfam\eighttt}
  \textfont\bffam=\eightbf  \scriptfont\bffam=\sixbf
    \scriptscriptfont\bffam=\fivebf  \def\bf{\fam\bffam\eightbf}
  \tt  \ttglue=.5em plus.25em minus.15em
  \normalbaselineskip=9pt
  \setbox\strutbox=\hbox{\vrule height7pt depth2pt width0pt}
  \let\sc=\sixrm  \let\big=\eightbig \normalbaselines\rm}

\font\eightrm=cmr8 \font\sixrm=cmr6 \font\fiverm=cmr5
\font\eighti=cmmi8  \font\sixi=cmmi6   \font\fivei=cmmi5
\font\eightsy=cmsy8  \font\sixsy=cmsy6 \font\fivesy=cmsy5
\font\eightit=cmti8  \font\eightsl=cmsl8  \font\eighttt=cmtt8
\font\eightbf=cmbx8  \font\sixbf=cmbx6 \font\fivebf=cmbx5

\def\eightbig#1{{\hbox{$\textfont0=\ninerm\textfont2=\ninesy
        \left#1\vbox to6.5pt{}\right.\enspace$}}}

\def\ninepoint{\def\rm{\fam0\ninerm}  
  \textfont0=\ninerm \scriptfont0=\sixrm \scriptscriptfont0=\fiverm
  \textfont1=\ninei  \scriptfont1=\sixi  \scriptscriptfont1=\fivei
  \textfont2=\ninesy  \scriptfont2=\sixsy  \scriptscriptfont2=\fivesy
  \textfont3=\tenex  \scriptfont3=\tenex  \scriptscriptfont3=\tenex
  \textfont\itfam=\nineit  \def\it{\fam\itfam\nineit}
  \textfont\slfam=\ninesl  \def\sl{\fam\slfam\ninesl}
  \textfont\ttfam=\ninett  \def\tt{\fam\ttfam\ninett}
  \textfont\bffam=\ninebf  \scriptfont\bffam=\sixbf
    \scriptscriptfont\bffam=\fivebf  \def\bf{\fam\bffam\ninebf}
  \tt  \ttglue=.5em plus.25em minus.15em
  \normalbaselineskip=11pt
  \setbox\strutbox=\hbox{\vrule height8pt depth3pt width0pt}
  \let\sc=\sevenrm  \let\big=\ninebig \normalbaselines\rm}

\font\ninerm=cmr9 \font\sixrm=cmr6 \font\fiverm=cmr5
\font\ninei=cmmi9  \font\sixi=cmmi6   \font\fivei=cmmi5
\font\ninesy=cmsy9  \font\sixsy=cmsy6 \font\fivesy=cmsy5
\font\nineit=cmti9  \font\ninesl=cmsl9  \font\ninett=cmtt9
\font\ninebf=cmbx9  \font\sixbf=cmbx6 \font\fivebf=cmbx5
\def\ninebig#1{{\hbox{$\textfont0=\tenrm\textfont2=\tensy
        \left#1\vbox to7.25pt{}\right.$}}}

\def\S{{\cal S}}
\def\Z{{\Bbb Z}}
\def\chix{{\raise.5ex\hbox{$\chi$}}}
\def\chixa{{\chix\lower.2em\hbox{$_A$}}}

\def\real{{\rm I\kern-.2em R}}
\def\integer{{\rm Z\kern-.32em Z}}
\def\complex{\kern.1em{\raise.47ex\hbox{
            $\scriptscriptstyle |$}}\kern-.40em{\rm C}}
\def\vs#1 {\vskip#1truein}
\def\hs#1 {\hskip#1truein}
  \hsize=6.2truein \hoffset=.23truein 
  \vsize=8.8truein 
\pageno=1 \baselineskip=12pt
  \parskip=0 pt \parindent=20pt 
\overfullrule=0pt \lineskip=0pt \lineskiplimit=0pt
  \hbadness=10000 \vbadness=10000 
     \pageno=0
     
     \footline{\ifnum\pageno=0\hss\else\hss\tenrm\folio\hss\fi}
     \hbox{}
     \vskip 1truein\centerline{{\bf First Order Phase Transition in a Model of Quasicrystals}}
     \vskip .2truein\centerline{by}
     \vskip .2truein
\centerline{{David Aristoff}
\ \ and\ \  {Charles Radin}
\footnote{*}{Research supported in part by NSF Grant DMS-0700120\hfil}}

\vskip .1truein
\centerline{ Mathematics Department, University of Texas, Austin, TX 78712} 
\vs.5 \centerline{{\bf Abstract}} 
\vs.2 \nd
We introduce a family of two-dimensional lattice models of
quasicrystals, using a range of square hard cores together with a soft
interaction based on an aperiodic tiling set. Along a low temperature
isotherm we find, by Monte Carlo simulation, a first order phase
transition between disordered and quasicrystalline phases.

\vs2
\centerline{February, 2011}
\vs1
\centerline{PACS Classification:\ \ 82.35.Lr, 64.70.M-, 64.60.De, 36.20.Fz}

     \vfill\eject
\nd
{\bf 1. Introduction}
\vs.1

Certain alloys have been shown to exhibit exotic 
structures in the sense that the structures produce sharp X-ray Bragg peaks 
with symmetries which
cannot arise from any crystal [1]. We model
these materials with equilibrium statistical mechanics
using a technique due to Levine and Steinhardt [2] which employs an 
energy ground state associated with an ``aperiodic tile set'', that is, a set of
several geometric shapes which only achieve their densest packings in
nonperiodic ways, the best known two-dimensional example being the
kite and dart shapes of Penrose [3]. There is very little direct
evidence that any equilibrium statistical mechanics model 
would show a well-defined quasicrystalline phase 
at positive temperature; we know only one model for
which there is (simulation) evidence [4,5] of such a phase. The theory of
solid/fluid phase transitions often emphasizes the role of crystal
symmetry, going back at least to Landau [6,7], and it is of interest to
understand how the quasicrystalline state, with its exotic form of
symmetry [8], fits into this picture. In
this regard it is noteworthy that the simulations in [4,5] indicate a
continuous (high order) transition between the fluid and quasicrystalline phases. We
revisit this issue here with slightly more realistic models and
find clear simulation evidence of a discontinuous (first order)
transition. We contrast our results with related models of hard
squares and of Widom-Rowlinson type.

\vs.2 \nd
{\bf 2. The Model}
\vs.1

Our models are variants of that used in [4,5], which we describe
first. The lattice model of [4,5] uses 16 types of particles, each
corresponding to an element of a certain subset $\S\subset \{1,2,3,4,5,6\}^4$ of
cardinality 16, associated with an aperiodic tiling set due to Ammann [9]. 
The model consists of configurations of particles on the square lattice 
$\Z^2$. It is convenient to interpret each element of $\S$ as a unit-edge
square {\it tile}, with its 4 edges colored
using 6 possible colors $\{1,2,3,4,5,6\}$ in one of the 16 groups of
4 indicated by $\S$; see Fig.\ 1. 
\vs0
\epsfig .9\hsize; 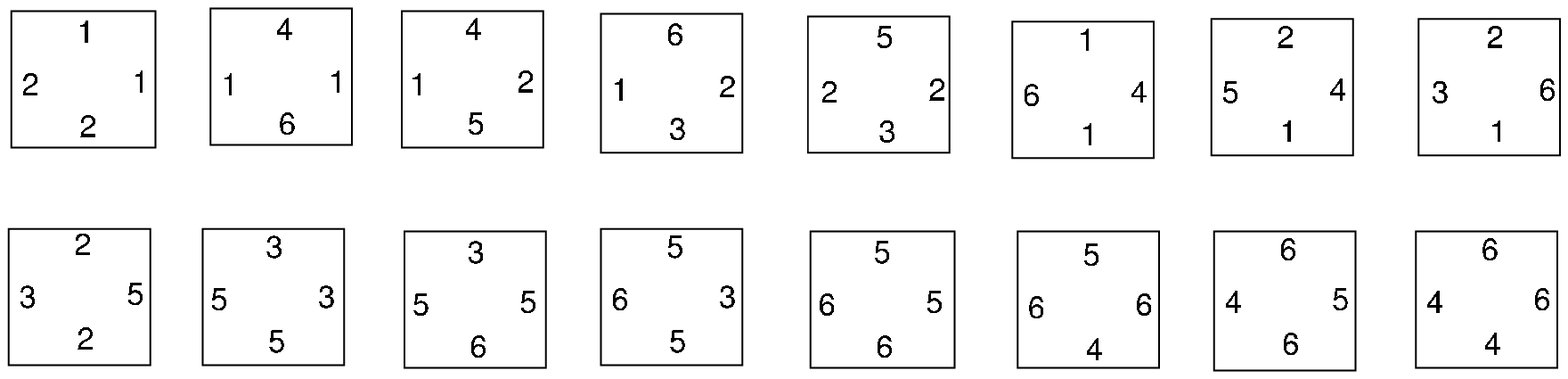;
\vs-.5
\centerline{Figure 1. Ammann's 16 Wang tiles.}
\vs.2
The model in [4,5] requires precisely
one particle/tile centered at each lattice site, and each particle interacts
with each of its 4 nearest neighbors only, with interaction energy $0$ or 
$-1$, with the negative value when the edge colors match. 
A canonical ensemble is used, but with density fixed at 1.
The model then exhibits a continuous phase transition at
positive temperature [4,5], with the low temperature phase showing
quasicrystalline structure, including the symmetry [8] it inherits from
Ammann's aperiodic tiling set.

We consider variants of the above model, investigating the
consequences of two changes: extending the size of the square tiles to
integer edge length $w\ge 1$ (they are still ``hard'' -- no overlap of tiles is
allowed); and
allowing a full range from 0 to 1 of volume fraction for the particle
configurations, in place of the fixed volume fraction of 1 used in
[4,5]. Note that the latter adds a second thermodynamic variable to the model.

The description of the interaction is a bit more complicated
than in the $w=1$ model of [4,5], since tiles which are close but not touching 
may still interact, though with lower amplitude. To define the
interaction we need some notation.
A {\it tile-state} $A$ is an element of $\Z^2 \times {\cal
S}$, with the first coordinate of $A$ representing its center on the
lattice, and the second coordinate representing its edge colors.
We define a pairwise interaction $H$ between tiles as follows. For
each tile-state $A$, consider the four ``nearest neighbor'' (empty)
squares of the same size as $A$ which surround $A$ but do not overlap it,
labelled $N$, $S$, $W$ and $E$ as in Fig.\ 2. If another tile-state $A'$ 
is disjoint with $A$ but

\vs-.5
\epsfig .9\hsize; 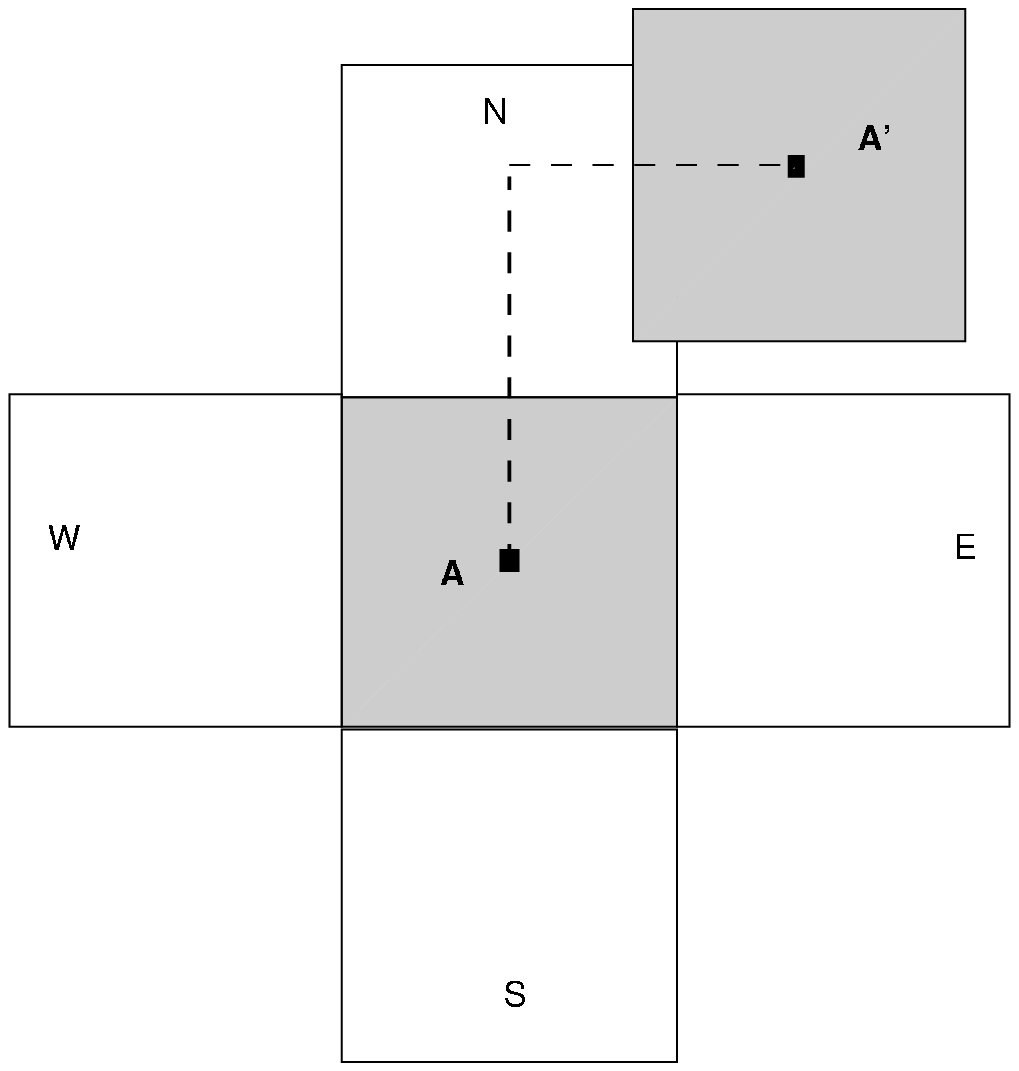;
\vs-.8
\centerline{Figure 2. The interaction energy depends on the }
\centerline{$\ell^1$ distance between particle centers (dashed line).}
\vs.2

\nd overlaps one of these four squares -- it cannot possibly overlap
more than one -- its energy of interaction with $A$, $H(A,A')$, is
as follows. Assume
without loss of generality that $A'$ overlaps $N$. Then $H(A,A')$:
is negative if and only if the north edge of $A$ has the same color as the south edge
of $A'$, and positive otherwise; and has absolute value given by 
$|H(A,A')| = (3w-d)/(2w)$, where $d$ is the $\ell^1$ distance between 
the centers of $A$ and $A'$. (The
$\ell^1$ distance between $(x,y)$ and $(x',y')$ is $|x-x'|+|y-y'|$.)
If $A'$ overlaps $S$, $W$, or $E$, then the energy $H$ is defined
similarly.

Given an inverse temperature $\beta$, a chemical potential 
$\mu$ (common to all tiles/particle types), 
and a finite volume ${\cal V} \subset \Z^2$, 
we consider the {grand canonical ensemble}: the 
probability $\cal P_{\beta,\mu}$ on configurations $C$ 
of tile-states in ${\cal V}$ which is given by 
$${\cal P_{\beta,\mu}}(C) = {\exp{[-\beta({\cal H}(C)-\mu {\cal
N}(C))]} \over Z},\eqno{(1)}$$ 
where ${\cal N}(C)$ is the number of tiles-states in $C$, 
${\cal H}(C) = \sum_{A\ne A' \in C} H(A,A')$ is the total energy 
of $C$, and $Z=Z(\mu,\beta)$ is the appropriate normalization.
In our simulations we
\vs0
\epsfig .6\hsize; 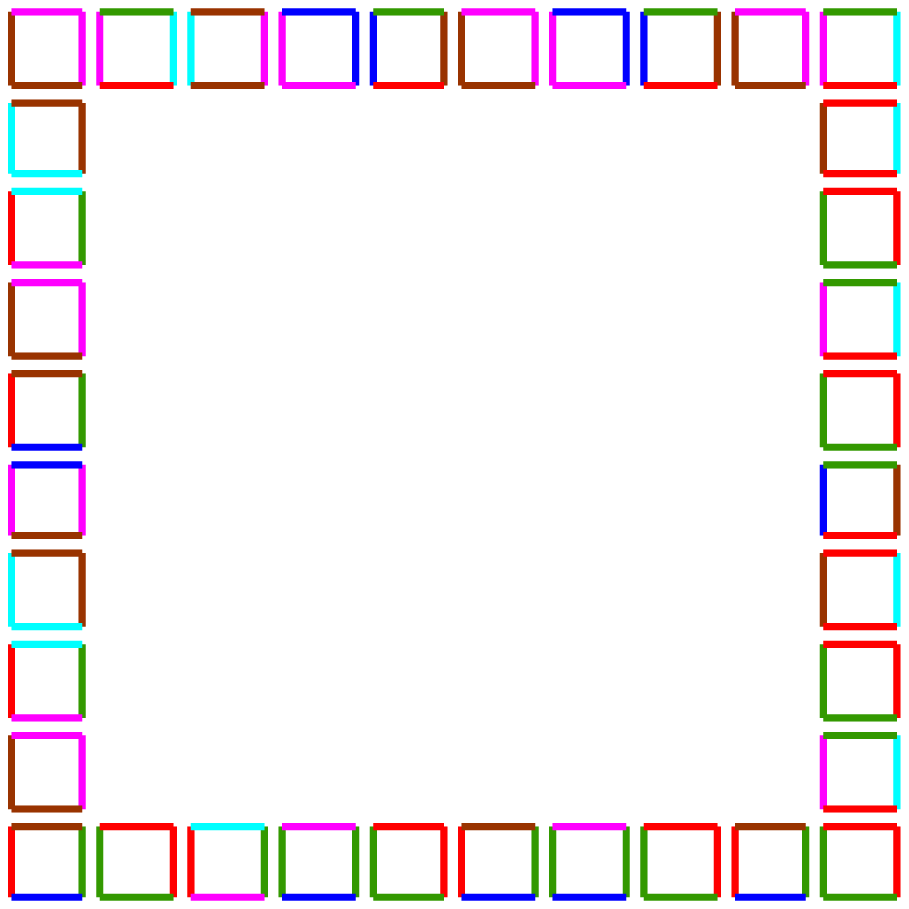;
\vs-.5
\centerline{Figure 3: A boundary shell (with edges colored instead of numbered).}
\vs.2

\nd use ``square'' volumes ${\cal V} = \Z^2 \cap [-k,k]^2$ with $k$ an integer 
multiple of $w$. We use boundary conditions as follows. Starting from a perfect 
tiling $\cal T$ by tile-states, that is, a configuration of tiles 
covering $\Z^2$ in which all edge colors match, we remove all the tiles 
except those just outside ${\cal V}$; this leaves a (fixed) boundary ``shell''; 
see Fig.\ 3. (We use 10,000 different boundary shells arising from random translations of 
a single tiling.)

It remains to define the order parameter, $\chi$. Each tile-state has
a ``long/short pattern'' given by its 4 edge colors, where we call colors
1 and 2 ``short'' and colors 3, 4, 5 and 6
``long''. Note that, for any tile, horizontal edges are either
both short or both long, and vertical edges are either both short or both long, 
so there are 4 possible long/short patterns for a tile. Any tile-state $A$ centered
``near'' the site $(wi,wj)\in {\cal V}$, namely centered in
$[wi-w/2,wi+w/2)\times[wj-w/2,wj+w/2)$, will be said to {\it agree
with} ${\cal T}$ if it has the same long/short pattern as
the tile-state in $\cal T$ centered at $(wi,wj)$.
Now, consider the thick ``shells'' 
${\cal V}_j = {\cal V} - [-j,j]^2$.
Given a configuration $C$ in ${\cal V}$ and boundary from $\cal T$, let $j^*$ be the largest 
integer such that at least $80\%$ of sites $(wi,wj)$ in ${\cal V}_{j^*}$ have a 
tile-state centered nearby (in the above sense) which agrees with $\cal T$. 
The order parameter $\chi$ is defined as the corresponding normalized volume, 
$\chi(C) = |{\cal V}_{j^*}|/|{\cal V}|$.
\vs.2 \nd
\nd {\bf 3. Simulations, and Results}
\vs.1
We ran Markov-chain Monte Carlo simulations of the model with tile widths 
$w=1$, $2$, and $3$. For each tile width $w$ we simulated 
volumes $|{\cal V}| = (20w)^2$, $(40w)^2$, $(60w)^2$, $(80w)^2$ and $(100w)^2$. 
(We will also use the notation $V = |{\cal V}|/w^2$, so $V$ represents 
the maximum possible number of tile-states in a single configuration.) 
The Monte Carlo steps were designed to produce the target distribution of equation (1). 
The basic Monte Carlo step involves a choice between four types of moves:
moving a particle (locally), changing a particle type, removing a particle, and 
adding a particle. As usual in grand canonical Monte Carlo simulation, given 
a configuration $C$ a trial configuration $C'$ is introduced, effecting one 
of the moves described above, and is accepted with probability determined by 
the desired limiting distribution, that is, equation (1). 

More precisely, the basic Monte Carlo step is as follows. 
Let $C(n)$ be the configuration at the $n$th step of the Monte Carlo chain, and  
choose a random lattice site $x$. Select $k \in \{1,2,3,4\}$ with probability $p_k$, 
where the $p_k$ are the probabilities of attempting the four types of 
Monte Carlo moves, with the $p_k$ summing to 1 and $p_3 = p_4$. 
(We choose the $p_k$ somewhat arbitrarily.) 
Without loss of generality suppose $x \in R:=[wi-w/2,wi+w/2)\times[wj-w/2,wj+w/2)$. 
First suppose $k\in \{1,2\}$. If there is no tile-state 
centered in $R$, then $C(n+1) = C(n)$; otherwise, let $A$ 
be the (unique) tile-state centered in $R$, 
and produce a trial configuration $C(n)'$ by: if $k=1$, moving the center of $A$ 
to a site in $R$ chosen uniformly at random; if $k=2$, replacing $A$ with a different 
tile-state centered at the same lattice site. Now suppose $k = 3$. If 
there is a tile-state $A$ centered in $R$, then $C(n)'$ is obtained by removing $A$; 
otherwise $C(n+1) = C(n)$. Finally suppose $k=4$. If there is no tile-state centered in $R$, 
then let $C(n)'$ be the configuration obtained from $C(n)$ by adding a tile-state chosen 
uniformly at random and centered at $x$; otherwise $C(n+1) = C(n)$. 
In cases where a trial configuration 
$C(n)'$ is introduced, we take $C(n+1) = C(n)'$ with probability $Q = \min(1,q)$; 
otherwise we take $C(n+1) = C(n)$. Here $q$ is given by 
$$q := \min\{1,e^{\beta[{\cal H}(C)-{\cal H}(C')+\mu^*({\cal N}(C')-{\cal N}(C))]}\}, \eqno{(2)}$$
where we have written $C$ for $C(n)$ and $C'$ for $C(n)'$, and take ${\cal H}(C') = \infty$ for 
trial configurations $C'$ with overlapping tiles. Here $\mu^* = \mu+\beta^{-1}\ln(16s^2)$ is 
chosen so that the Monte Carlo configurations $C(n)$ have limiting probability 
distribution given by equation (1). (The value $\mu^*$ arises because of the following. 
Starting with a  
configuration $C$, the probability of proposing a trial configuration $C'$ which 
removes tile-state $A$ from $C$ is $16s^2$ times the probability that, 
starting with the configuration $C'$, a trial configuration $C''=C$ is 
proposed which adds 
the tile-state $A$ to $C'$.)

For $w=2$ and $w=3$ we find clear evidence of discontinuous (first order) phase 
\vfill \eject

\vs0
\epsfig 1.0\hsize; 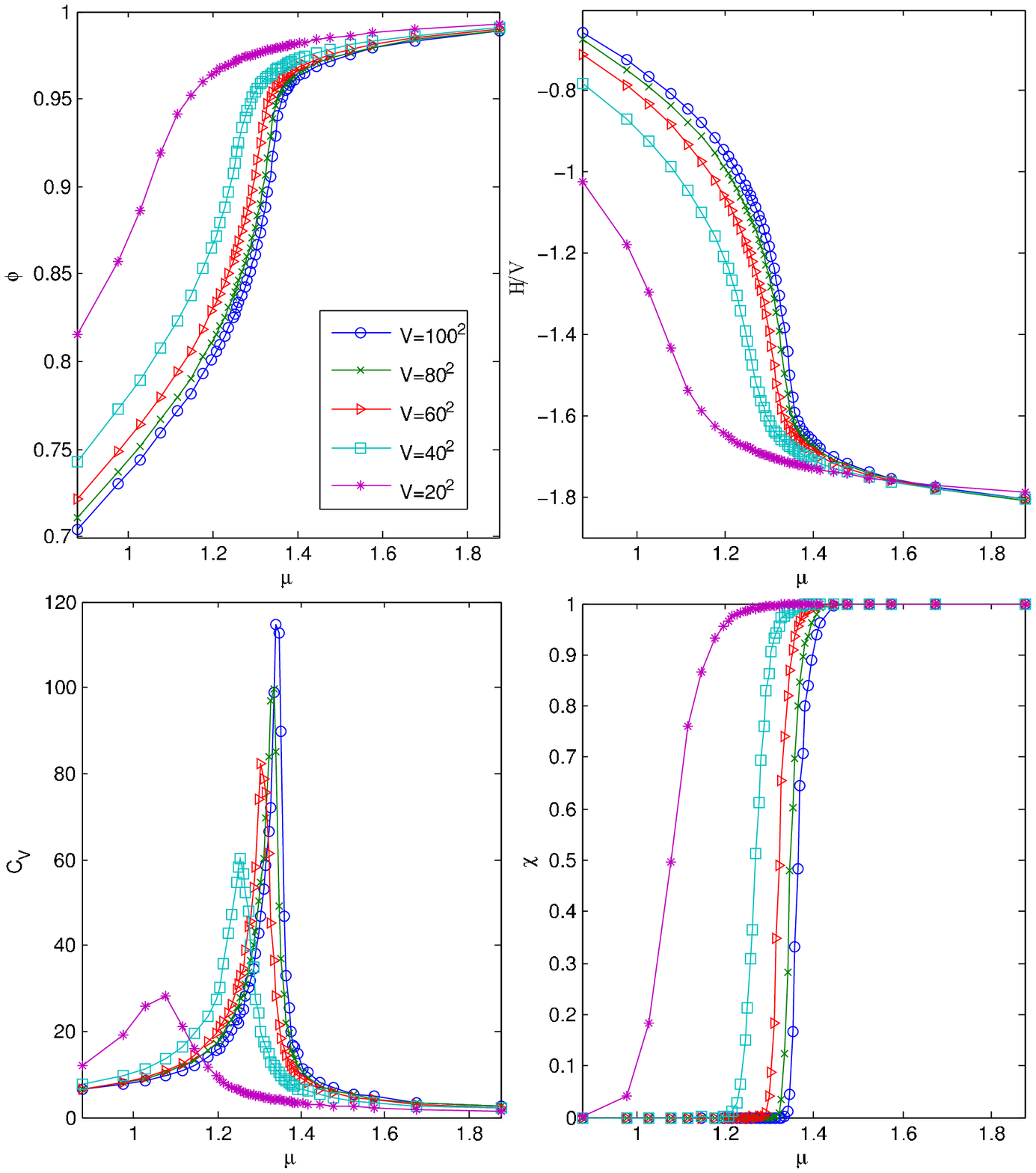;
\vs0
\centerline{Figure 4: Data for $w=2$ and $\beta = 1.5$.}
\vs.2
\nd
transitions. At the transitions 
(at $\beta=1.5$, near $\mu = 1.4$ for $w=2$ and near $\mu = 2.0$ for $w=3$)
the average volume fraction 
$\phi$ and energy per volume ${\cal H}/V$ exhibit developing jump discontinuities 
as $V$ increases; see Figs. 4--5. 
Furthermore the (volumetric) heat capacity, $C_V$ (defined as the 
derivative of ${\cal H}/V$ with respect to $T = 1/\beta$), exhibits a developing 
delta function as $V$ increases (see Figs. 4--5). 
\vs0
\epsfig 1.0\hsize; 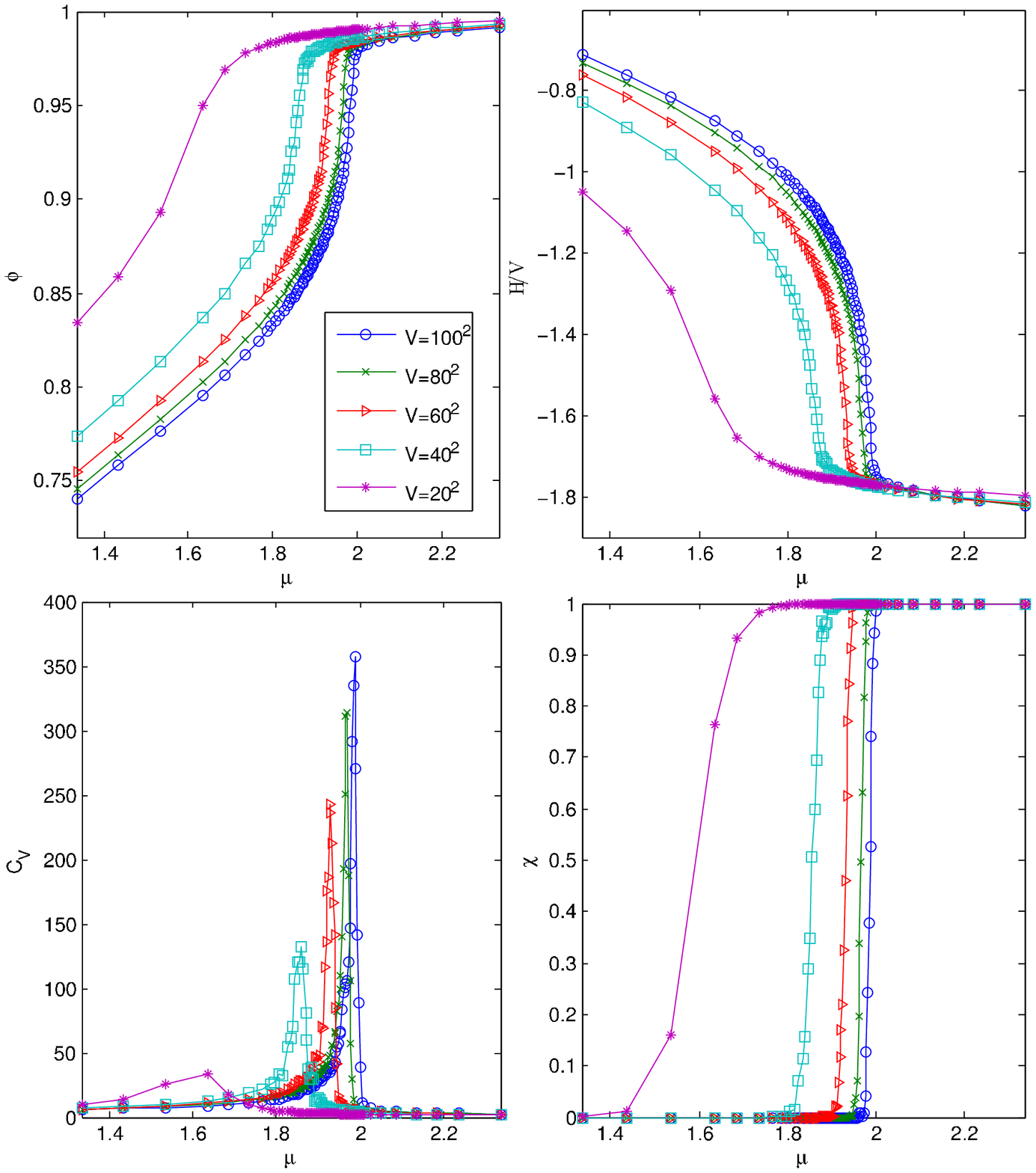;
\vs0
\centerline{Figure 5: Data for $w=3$ and $\beta = 1.5$.}
\vs.2
 From our data in Fig.\ 6, for the $w=1$ model, there appears to be a
continuous transition (at $\beta=1.5$ near $\mu=0.8$), from a disordered fluid phase to
a phase with quasicrystalline order, similar to the transition in [4,5] of 
the canonical ensemble at density 1.
\vfill \eject

\vs.2
\epsfig 1.0\hsize; 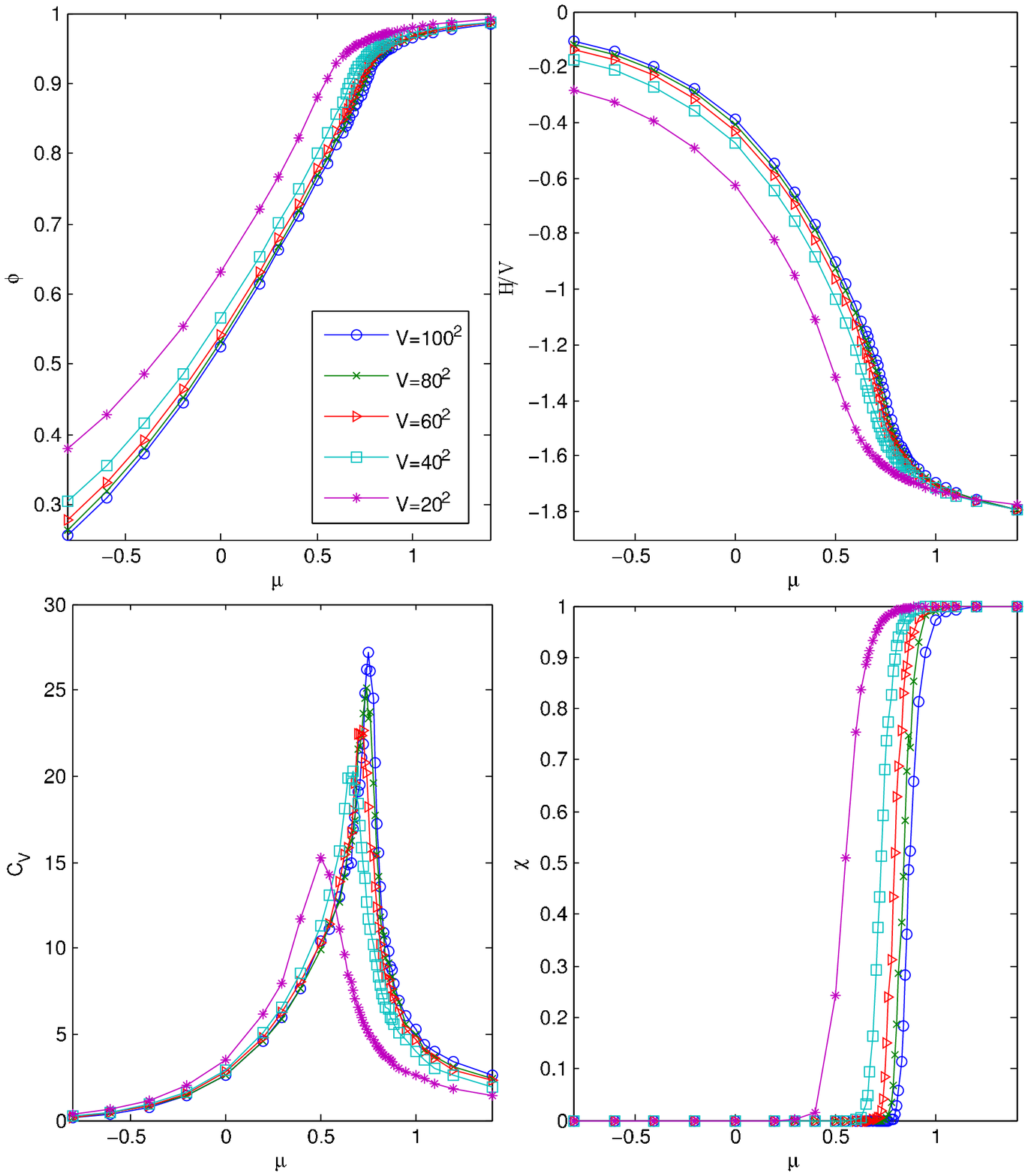;
\vs-.2
\centerline{Figure 6: Data for $w=1$ and $\beta = 1.5$.}
\vs.2

To quantify the developing delta function in $C_V$ noted above for
$w=2,3$ we  
measure, from Figs. 4 and 5, the maximum value $h$ of $C_V$ divided by 
the ``width'' of the graph of $C_V$ at $h/2$ (see Fig.\ 7). 
The order parameter $\chi$ also develops a discontinuity at the transition, 
signaling the onset of quasicrystalline symmetry: see Figs. 4--5. 
\vs0
\epsfig 1.0\hsize; 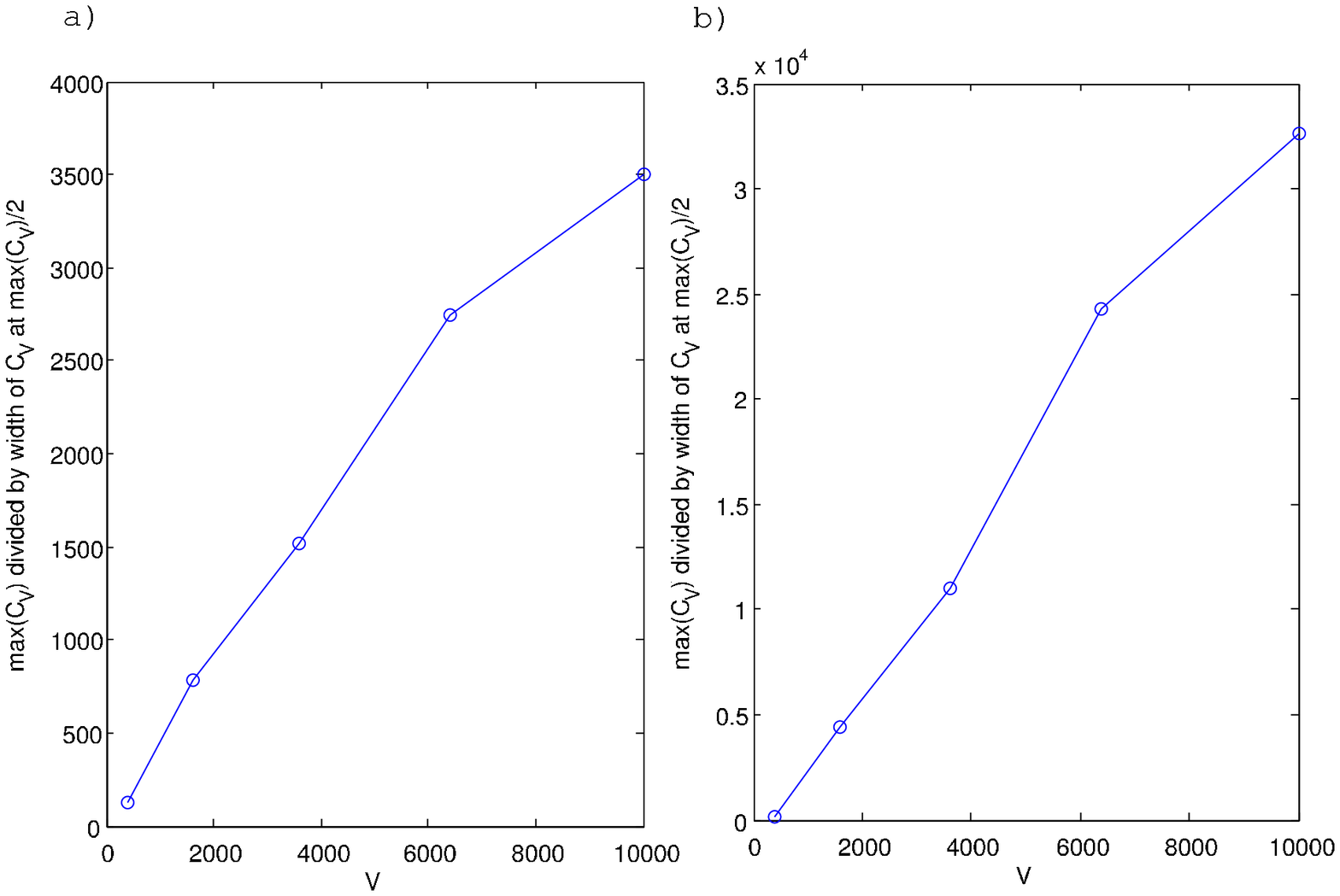;
\vs0
\centerline{Figure 7. Evidence of $C_V$ developing into a delta function for a) $w=2$, and b) $w=3$.}
\vs.2
\nd {\bf 4. Comparisons} 

\vs.1 We next compare our results with related models, in particular
hard squares and Widom-Rowlinson models on $\Z^2$; see [10] for a
review of, and earlier references to, the former and [11] for a review
of, and earlier references to, the latter.

At sufficiently high temperature the soft interaction between 
our particles must
become negligible compared to the hard core so we first review what is
known about hard square systems.

For hard squares on $\Z^2$ the energy is zero for all configurations
so it is convenient to use a grand canonical ensemble with fixed
temperature 1 and variable chemical potential $\mu$.  Hard squares
with edge size $w=1$ (``point hard core'') do not have a phase transition as $\mu$ is
varied, since the variables at each site, `occupied' or `unoccupied',
are independent. For fixed $w\ge 2$ the densest configurations, of
volume fraction 1 and corresponding to $\mu\to \infty$, are
degenerate, allowing parallel rows or columns of squares to slide
independently in what is called a ``columnar'' phase. It is generally
accepted from simulation [10] that the model with $w=2$ has a continuous
transition at volume fraction about $0.92$. We only know of one paper 
on the model with $w=3$ [10]; the authors claim a first order freezing transition 
beginning at a density above $0.91$, but are unable to show developing 
discontinuities in any thermodynamic quantities. 

We now compare our model to hard squares on $\Z^2$. 
Consider a version of our quasicrystal model with $w \ge 2$ but in a
canonical ensemble with 
volume fraction fixed at 1. Based on the results of [4,5], we expect this model 
to have a columnar phase at high temperature, and a quasicrystalline 
phase at 
low temperature, with a transition between
them at some unknown 
temperature $T_c$. 
Now considering the same model except at lower volume fraction or
chemical potential, one expects the
transition to the quasicrystalline phase to 
move to lower temperature since the soft interaction should have less effect
as the density is lowered. We conclude that above temperature $T_c$ our model 
should behave 
qualitatively like hard squares, with disordered and columnar phases,
but without a phase with quasicrystalline order.

Absent from the above discussion is the effect on entropy, in our model, 
of the multiple types of particle. This effect is emphasized in Widom-Rowlinson models, 
which we now discuss. The multitype Widom-Rowlinson model with $q$ colors, on $\Z^2$, is 
defined by the hard core condition that nearest neighbor sites cannot be
occupied by different colors. The main conclusion of such work is the
existence (for large enough $q$) of two types of high density phase
transitions: one transition at which spatial symmetry is broken,
producing different population densities for the two sublattices, but without
breaking the symmetry between particle types, and a second
(discontinuous) transition at higher density at which the symmetry
between particle types is broken, so that one particle type dominates
the density [11]. These models, with diamond-shaped instead of square hard cores, 
are not simply related to our models, but there are two interesting
variants treated by Georgii and Zagrebnov [12] which are closely related. 
The first model has a square-shaped hard core, with square width $w=2$, between 
particles of different
color; the second model adds to this a smaller diamond-shaped hard 
core between nearest neighbor particles of the same color. One could 
think of these models as employing a hard core of size $w=2$ between {\it all}
particles, with a ``cancelling'' (or attractive) interaction between
particles of like color on sites at separation $\sqrt{2}$ for the
second model, and for sites at separation $1$ and $\sqrt{2}$ for the
first model. Thus the models can be thought of as hard squares with 
an added particle-type-dependent short range attraction, somewhat like our
quasicrystal models. Both Widom-Rowlinson models show only one
transition, which is discontinuous. For the first model the high density 
phase breaks particle type symmetry but not spatial symmetry, and for the second model 
the high density phase breaks the symmetry of both space and particle type. 
In particular the second model seems close to our ($w=2$) quasicrystal
model, and our results bear this out, with a discontinuous transition
corresponding to broken spatial and particle-type symmetries. 
The correspondence might be closer if the attraction and repulsion in
our model was a hard rather than soft interaction, as in [13].

It would have been useful to simulate our model over a range of
temperatures, but we found this prohibitively expensive in computation
time. For this reason we investigated a technique noted in [14], in
which the initial state for the simulation is an energy ground
state. However this technique proved to be unreliable here unless the
runs were of comparable length to those starting from vacuum, thus
yielding no advantage (see Fig.\ 8). (We suspect that simulations starting
from an energy ground state get restricted to a narrow portion of
phase space more easily than those starting from the vacuum.)

\vs0
\epsfig .7\hsize; 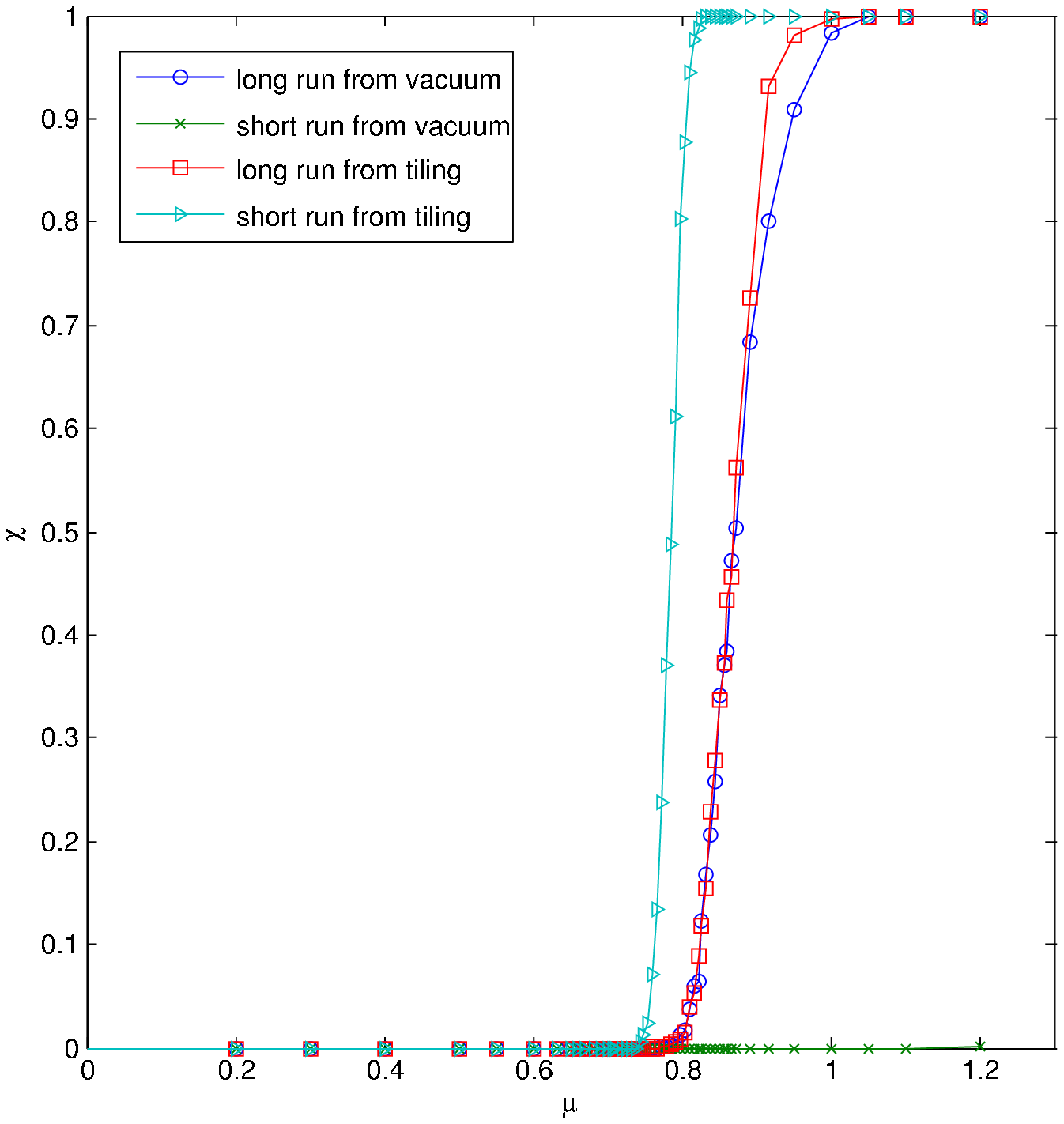;
\vs-.2
\nd Figure 8. Comparison of order parameter values of long and short runs starting from 
{both vacuum and perfect tiling, for $w=1$ and $V=100^2$. (The long runs are 100 times longer than the short runs.)}

\vs.2 \nd
{\bf 5. Conclusion}
\vs.2

We have introduced an extension of a two dimensional lattice gas model of a
quasicrystal [4,5]. That model was studied as a function of temperature at fixed 
density 1, with simulations showing a continuous phase transition between a disordered 
(fluid) phase and a phase with quasicrystalline order. When we
introduce vacancies
into the model we find that the basic character does not seem to
change: there still appears to be only the two phases, disordered and
quasicrystalline, with a continuous transition between them.

We then generalized the model by extending the range of the hard core
as well as allowing variation in the density, and found, by
simulation along a (low temperature) isotherm, that the model
then exhibits a {\it discontinuous} transition between a low density
disordered and high density quasicrystalline phase. The high density 
quasicrystalline phase breaks symmetries of space (a square sublattice 
is preferentially populated) and particle type (the particles inherit 
the quasicrystalline boundary structure). Along an isotherm
at sufficiently high temperature we expect the model to exhibit the
transition of simple hard squares (continuous for $w=2$ and possibly
discontinuous for $w=3$), between the low density disordered and high
density columnar phases. At fixed high density or chemical potential
we expect a transition from the quasicrystalline phase to a columnar
phase as temperature is raised from zero. 

This family of models with unusual symmetry adds a special flavor to 
the more traditional lattice gas models with hard core. Its
complicated multiparticle structure seems at present to defy proof
of a transition, and is also expensive to simulate. Further progress in
filling out the phase structure of the model would be highly
desirable.


\vs.3

\centerline{{\bf References}}
\vs.2
\item{1)} D. Shechtman, I. Blech, D. Gratias and J.W. Cahn,
Metallic phase with long-ranged orientational order and no
tranlational symmetry,
Phys. Rev. Lett. 53 (1984) 1951-53.

\item{2)} D. Levine and P.J. Steinhardt, 
{Quasicrystals: a new class of ordered structures},
Phys. Rev. Lett. {53} (1984) 2477-2480.

\item{3)} M.\ Gardner, Extraordinary nonperiodic tiling that enriches the
theory of tiles, {Sci.\ Am.\ (USA)} 236 (1977) 110-119.

\item{4)} L.~Leuzzi, G.~Parisi,
{Thermodynamics of a tiling model},
J. Phys. A {33} (2000) 4215--4225.

\item{5)} H. Koch and C. Radin, Modelling quasicrystals at positive temperature,
J. Stat. Phys. 138 (2010) 465-475.

\item{6)} L.D. Landau and E.M. Lifshitz, Statistical Physics,
Pergamon Press, London 1958, trans. E. Peierls and R.F. Peierls, Chapter XIV.

\item{7)} P. W. Anderson, Basic Notions of Condensed Matter Physics 
Benjamin/Cummings, Menlo Park, 1984, chapter 2.

\item{8)} C. Radin, Symmetries of quasicrystals, J. Stat. Phys. 95 (1999), 827-833.

\item{9)} B.\ Gr\"unbaum and G.C.\ Shephard, {Tilings and Patterns},
Freeman, New York, 1986, page 593.

\item{10)} H.C.M. Fernandes, J.J. Arenson and Y. Levin,
Monte Carlo simulations of two-dimensional hard core lattice gases,
J. Chem. Phys. 126 (2007) 114508.

\item{11)} J.L. Lebowitz, A. Mazel, P. Nielaba and L. Samaj,
Ordering and demixing transition in multicomponent Widom-Rowlinson
models, Phys. Rev. E 52 (1995) 5985-5996.

\item{12)} H.-O. Georgii and V. Zagrebnov,
Entropy-driven phase transition in multitype lattice gas models,
J. Stat. Phys. 102 (2001) 35-67.

\item{13)} D. Ruelle, {Thermodynamic Formalism}, Addison-Wesley, New
York, 1978, section 4.15.

\item{14)} Z. Rotman and E. Eisenberg,
A finite-temperature liquid-quasicrystal transition in a lattice model,
arXiv:1009.1156v1.

\vfill
\end